\documentclass[%
mathleft,%
]{an}
\usepackage{graphicx}
\usepackage{times}
\usepackage{color}
\begin{document}

\Pagespan{1}{}
\Yearpublication{}%
\Yearsubmission{}%
\Month{}%
\Volume{}%
\Issue{}%

\title{
\begin{changed}
Statistical Analysis for the Q-factor of twin kHz QPOs
\end{changed}
}

\author{J. Wang\inst{1}\fnmsep\thanks{Corresponding author:
  joanwangj@126.com}, H. K. Chang\inst{1}, C. M. Zhang\inst{2}, D. H. Wang\inst{3,2} \and L. Chen\inst{3}}

\titlerunning{Statistical Analysis for the Q-factor of twin kHz QPOs}
\authorrunning{J. Wang et al}
\institute{
Institute of Astronomy, National Tsing Hua University, Hsinchu, 30013, Taiwan
\and
National Astronomical Observatories, Chinese Academy of Sciences, Beijing 100012, P. R. China
\and
Astronomy Department, Beijing Normal University, Beijing, 100875, China
}

\received{XXXX}
\accepted{XXXX}
\publonline{XXXX}


\abstract{Using the recently published data of twin kHz quasi-period
oscillations (QPOs) in neutron star low-mass X-ray binaries (LMXBs),
we study the different profiles between bright Z sources and less
luminous Atoll sources
\begin{changed} in a statistical way. We find the
quality factors of upper kHz QPOs show a narrow distribution both
for Z sources and Atoll sources, which concentrate at 7.98 and 9.75
respectively,  the quality factors of lower kHz QPOs show a narrow
distribution for Z sources and a broader distribution for Atoll
sources, which concentrate at 5.25 and 86.22 respectively.
\end{changed}
In order to investigate the
\begin{changed}
relation
\end{changed}
between
\begin{changed}
the
\end{changed}
quality factor
and
\begin{changed}
the peak
\end{changed}
frequency of kHz QPOs,
\begin{changed}
we fit the data with
power-law, linear and exponential functions, respectively.
There is an obvious trend that the quality factors increase with
the peak frequencies both for upper and lower QPOs.
\end{changed}
The implications of our results are discussed.}

\maketitle

\section{Introduction}
A number of neutron stars in low mass X-ray binaries show the
kilohertz quasi-periodic oscillations
\begin{changed}
in
\end{changed}
their X-ray spectra
(van der Klis 2000, 2006). These frequencies, in the range of $200
\sim 1300$ Hz, are
\begin{changed}
as
\end{changed}
the same order as the dynamical
\begin{changed}
timescale
\end{changed}
of the innermost
\begin{changed}
region
\end{changed}
of the accretion flow
around the stellar mass compact objects (van der Klis 2006, 2008).
Owing to the expected links with the orbital motion, most
\begin{changed}
work
\end{changed}
about kHz QPOs focus on the explanation for the nature of these
signals (e.g. Miller, Lamb \& Psaltis 1998; Stella \& Vietri 1998,
1999; Kluzniak \& Abramowicz 2001; Abramowicz et al. 2003; Zhang
2004). The kHz QPOs often
\begin{changed}
arise
\end{changed}
as simultaneous twin peaks
(upper $\nu_2$ and lower $\nu_1$ frequency) with frequencies
changing over time. These frequencies behave in a rather regular way
and follow the tight correlations between their frequencies and
other observed characteristic frequencies (see, e.g. Psaltis et al.
1998, 1999a; Psaltis, Belloni \& van der Klis 1999b; Stella, Vietri
\& Morsink 1999; Belloni, Psaltis \& van der Klis 2002; Titarchuk \&
Wood 2002; M¡äendez \& van der Klis 1999, 2000; M\'{e}ndez et al.
2001; Yu, van der Klis, Jonker 2001; Yu, van der Klis 2002).
Moreover, the correlation between the upper frequency and
\begin{changed}
the
\end{changed}
lower frequency across different sources can be roughly fitted by a
power-law function (see e.g. Psaltis et al. 1998, 1999a; Zhang et
al. 2006), and also by a linear model (see Belloni, M\'{e}ndez \&
Homan 2005, 2007).

The kHz QPOs in LMXBs are narrow features (peaks) in their power
density spectra (PDS), whose profiles can be described by the
Lorentzian function $P_{\nu}\propto A_{0} w/[(\nu-\nu_0)^2+(w/2)^2]$
($\nu_0$ is the peak frequency, $w$ is the full width at
half-maximum (FWHM), and $A_0$ is the amplitude of
\begin{changed}
the
\end{changed}
signal). The
ratio of peak frequency to FWHM is the quality factor,
\begin{equation}
Q \equiv \frac{\nu_0}{w}.
\end{equation}
Therefore, the kHz QPOs in neutron star LMXBs can be characterized
by three characteristic quantities, i.e. centroid frequency (i.e.,
peak frequency $\nu_0$), quality factor (Q $\equiv \nu_0$ / FWHM)
and fractional root-mean-squared (rms). The quality factor
characterizes the coherence of the signal, while the rms represents
a measure of the signal strength and
\begin{changed}
it
\end{changed}
is proportional to the square
root of the peak power contribution to the PDS. Each kHz QPO
corresponds to its quality factor (lower $Q_1$ and upper $Q_2$
quality factor) and centroid frequency (upper $\nu_2$ and lower
$\nu_1$ frequency).

In the past several  years, the large Rossi X-ray Timing Explorer
(RXTE) archive makes
\begin{changed}
it
\end{changed}
possible
\begin{changed}
to study this quantity
systematically in several sources.
\end{changed}
Using
\begin{changed}
the
\end{changed}
data from RXTE,
Barret et al. (2005a) studied
\begin{changed}
the source
\end{changed}
4U 1608-52 and
revealed a positive correlation between
\begin{changed}
the
\end{changed}
lower frequency
\begin{changed}
QPOs
\end{changed}
and its quality factors, up to a maximum of about $Q \sim
200$. Motivated by this idea, Barret, Olive \& Miller (2005b, 2006)
studied the QPO properties and the dependency of
\begin{changed}
its
\end{changed}
quality factor on
\begin{changed}
the peak
\end{changed}
frequency in source 4U 1636-536. It shows
that
\begin{changed}
the
\end{changed}
quality factors for lower and upper kHz QPOs of 4U
1636-536 follow different tracks in a
\begin{changed}
$Q~vs.~\nu$
\end{changed}
plot, i.e.
\begin{changed}
the
\end{changed}
quality factors
\begin{changed}
of
\end{changed}
lower kHz
\begin{changed}
increase
\end{changed}
with
\begin{changed}
its peak
\end{changed}
frequency up to
850 Hz (Q $\sim$ 200) and
\begin{changed}
drop
\end{changed}
precipitously to the highest
\begin{changed}
detectable frequency
\end{changed}
$\sim$ 920 Hz (Q $\sim$ 50), while
\begin{changed}
the quality factors
\end{changed}
of
upper kHz
\begin{changed}
QPOs increase
\end{changed}
steadily all the way to the highest
detectable frequency. Moreover,
\begin{changed}
the quality factors
\end{changed}
of lower
\begin{changed}
QPOs
\end{changed}
is higher than
\begin{changed}
those
\end{changed}
of upper
\begin{changed}
ones
\end{changed}
(Barret, Olive \& Miller
2005b,c; 2006). The rough similarity
\begin{changed}
was also
\end{changed}
extended to sources 4U
1735-44
\begin{changed}
and
\end{changed}
4U 1728-34 (Barret, Olive \& Miller 2006; Boutelier, Barret
\& Miller 2009; M$\acute{e}$ndez 2006; T$\ddot{o}$r$\ddot{o}$k
2009).

In this paper, we analyze the
\begin{changed}
distributions
\end{changed}
of
\begin{changed}
the
\end{changed}
quality factors for upper and lower kHz QPOs
\begin{changed}
in
\end{changed}
ten sources
--- five Atoll sources (namely 4U 1608-52, 4U 1636-53, 4U 1728-34,
4U 1820-30 and 4U 1735-44) and five Z sources (namely Sco X-1, Cyg
X-2, GX 17+2, GX 5-1 and GX 340+0). In order to investigate the
\begin{changed}
relation
\end{changed}
between
\begin{changed}
the quality factors
\end{changed}
and
\begin{changed}
its
\end{changed}
centroid
frequencies, we fit
\begin{changed}
the data with
\end{changed}
the three
\begin{changed}
functions
\end{changed}
, i.e. linear relation ($Q=a+b\nu/1000$), power-law relation
($Q=a(\nu/1000)^b$) and exponential relation ($Q=a\times
exp(b\nu/1000)$),
\begin{changed}
according
\end{changed}
to $Q-\nu$ tracks. We benefit
from the existing studies and use the published data from the
collection by M\'{e}ndez (2006, and reference therein). In section
2, we analyze the data for these sources and execute fitting for the
different source categories. Conclusions and discussions are
contained in section 3.

\section{Statistical Analysis for the Coherence of kHz QPOs}
In this part, we investigate the data statistically and choose
linear ($Q = a+b\nu/1000$), power-law ($Q = a(\nu/1000)^b$) and
exponential relation ($Q = a\times exp(b\nu/1000)$) for fitting,
where $a$ and $b$ are the undefined parameters.
\begin{changed}
Firstly, we
\end{changed}
put
\begin{changed}
the sources
\end{changed}
together and fit
\begin{changed}
them with
\end{changed}
above three
relations, then
\begin{changed}
we
\end{changed}
divide them into Atoll and Z sources and
\begin{changed}
execute
\end{changed}
the same fitting.

\subsection{Statistically Analysis for the Data}

\begin{figure}
\includegraphics[width=9cm]{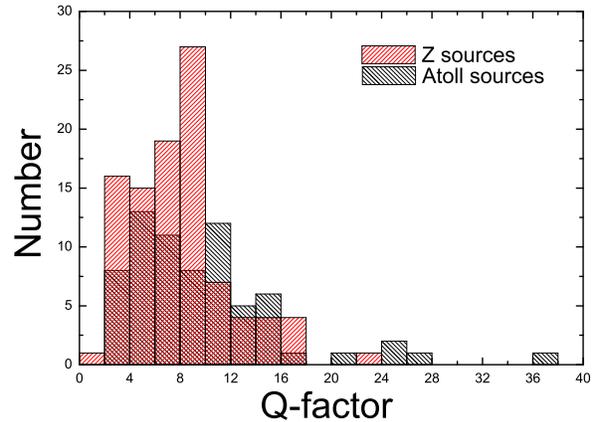}
\caption{The distributions of $Q_2$ for Z and Atoll sources. The red
shadow represents the $Q_2$ distribution for Z sources, while the
black shadow show that for Atoll sources.
{\color{red}
The mean value of $Q_2$ is $7.98$ for Atoll sources and $9.75$ for Z sources.
}
For the plot, we take the
width of bin as 2. \label{Fig:upper-his}}
\end{figure}

Firstly, we study the distributions of upper and lower quality
factors for Z and Atoll sources. Fig. \ref{Fig:upper-his} presents
the
\begin{changed}
distributions
\end{changed}
of upper quality factors ($Q_2$) for Z and Atoll
sources. It is found that most $Q_2$
\begin{changed}
of
\end{changed}
these two classes locate in
the
\begin{changed}
similar region
\end{changed}
(i.e. $Q_2$ = 2 - 18) as a whole. For the
exception, a few $Q_2$ of Atoll sources present relatively
\begin{changed}
larger
\end{changed}
values (up to 40, see Fig. \ref{Fig:upper-his} for detail). Most
$Q_2$ for Z sources gather in a range from 2 to 10
\begin{changed}
which
\end{changed}
is
larger than
\begin{changed}
those of
\end{changed}
Atoll sources. For a quantitative knowledge, we
calculate the mean value of $Q_2$ for Z and Atoll sources, i.e.
$<Q_{2Atoll}> = 9.75$ and $<Q_{2Z}> = 7.98$. The
mean value of $Q_2$ for Z sources is lower than that for Atoll
sources.

\begin{figure*}
\includegraphics[width=8.8cm]{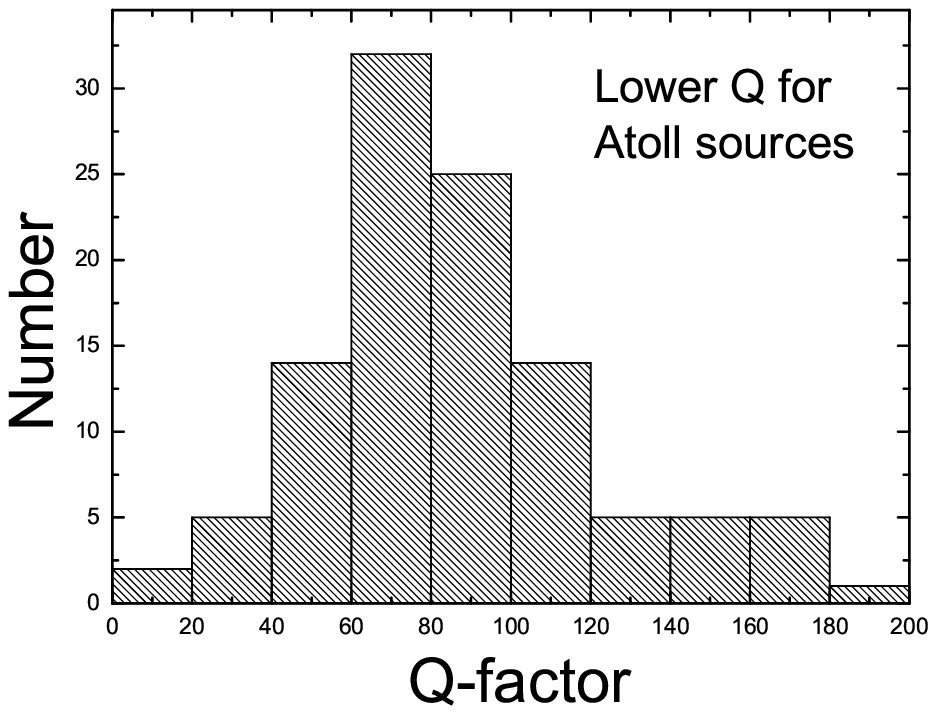}%
\includegraphics[width=8.34cm]{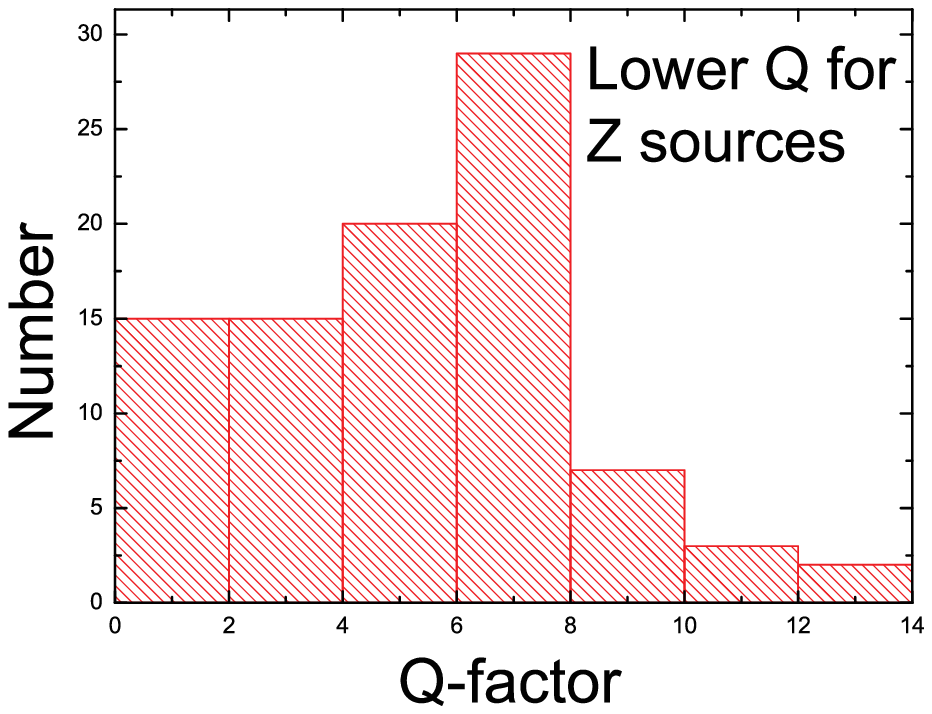}
\caption{The distributions of $Q_1$ for Z and Atoll sources. The
left panel is for Atoll sources, and the right one is for Z sources.
{\color{red}
The mean value of $Q_1$ is $86.22$  for Atoll sources and $5.25$ for Z sources.
}
For the plot, the width of bins are set to 20 and 2 for Atoll and Z
sources, respectively. \label{Fig:lower-his}}
\end{figure*}

In Fig. \ref{Fig:lower-his}, we plot the distributions of lower
quality factors ($Q_1$) for Z and Atoll sources, where we notice
very big difference in the range of $Q_1$ between
\begin{changed}
these two
\end{changed}
sources. The $Q_1$ for Z sources are
\begin{changed}
small
\end{changed}
($Q_{1Z} \leq 14$), and
\begin{changed}
those
\end{changed}
for Atoll sources are very
\begin{changed}
large
\end{changed}
(up to $Q_1 = 200$). The $Q_1$
ranges from 2 to 14 for Z sources and from 2 to 200 for Atoll
sources (see Fig. \ref{Fig:lower-his} for detail). Most data of
$Q_1$ for Z sources are distributed from 6 to 10, and that of Atoll
sources gather in the region of $Q_1 \sim 60-100$. Some $Q_1$ for
Atoll sources are as high as 200. We also calculate the mean values
of $Q_1$ both for Z and Atoll sources, i.e.
$<Q_{1Atoll}> = 86.22$, $<Q_{1Z}> = 5.25$.

On the whole, the quality factors for Z sources are
\begin{changed}
smaller
\end{changed}
than those
for Atoll sources. In addition, $Q_2$ for Z sources cover a
wider range than $Q_1$, while $Q_2$ cover the narrower range
than $Q_1$ for Atoll sources. The mean values
\begin{changed}
of
\end{changed}
$Q_1$ and $Q_2$
are $<Q_1> = 49.20$ and
\begin{changed}
$<Q_2> = 8.70$
\end{changed}
, respectively.

\subsection{Fitting for the $Q_2-\nu_2$ Relations}

\begin{figure*}
\includegraphics[width=5.65cm]{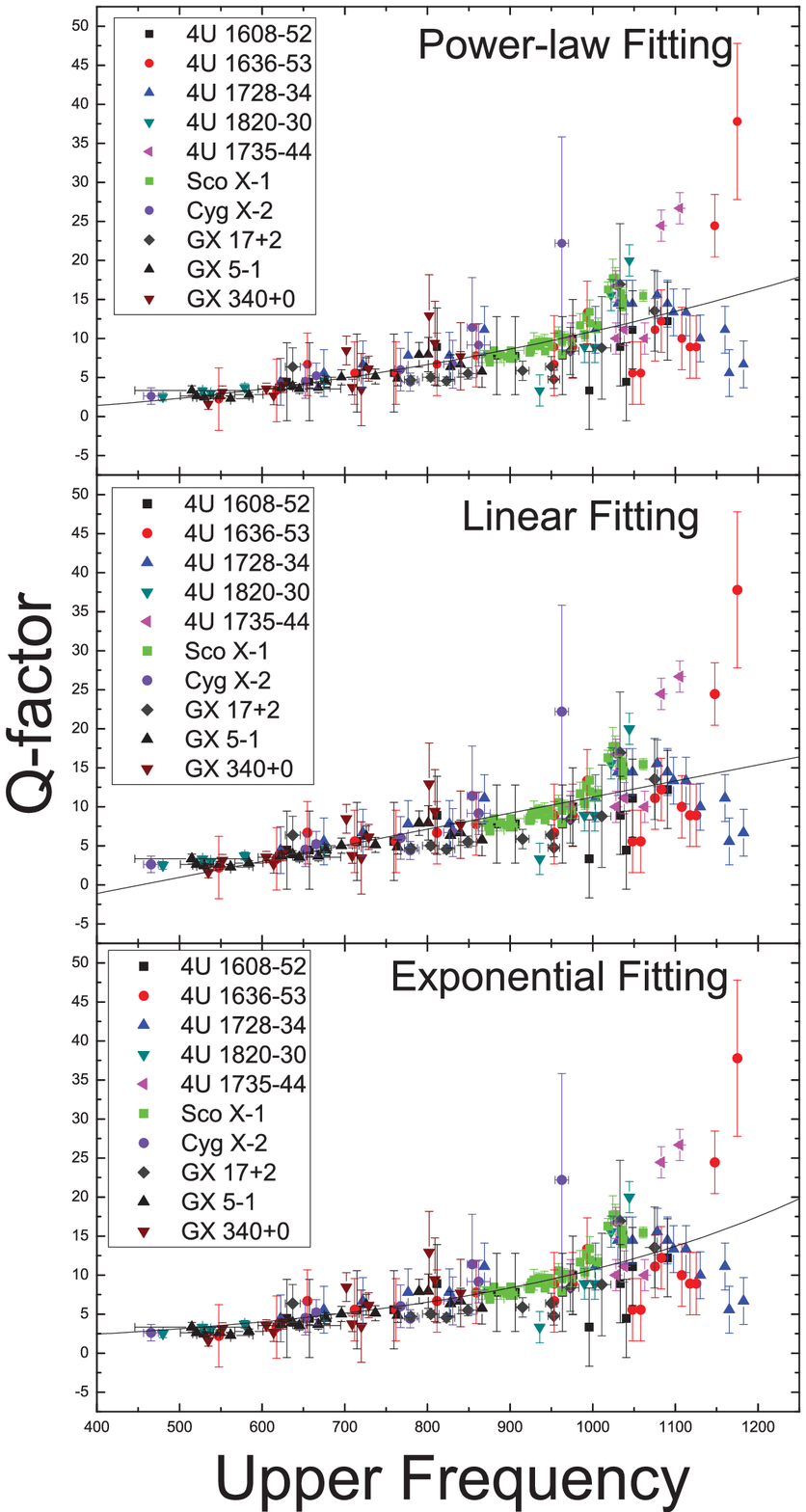}%
\includegraphics[width=5.76cm]{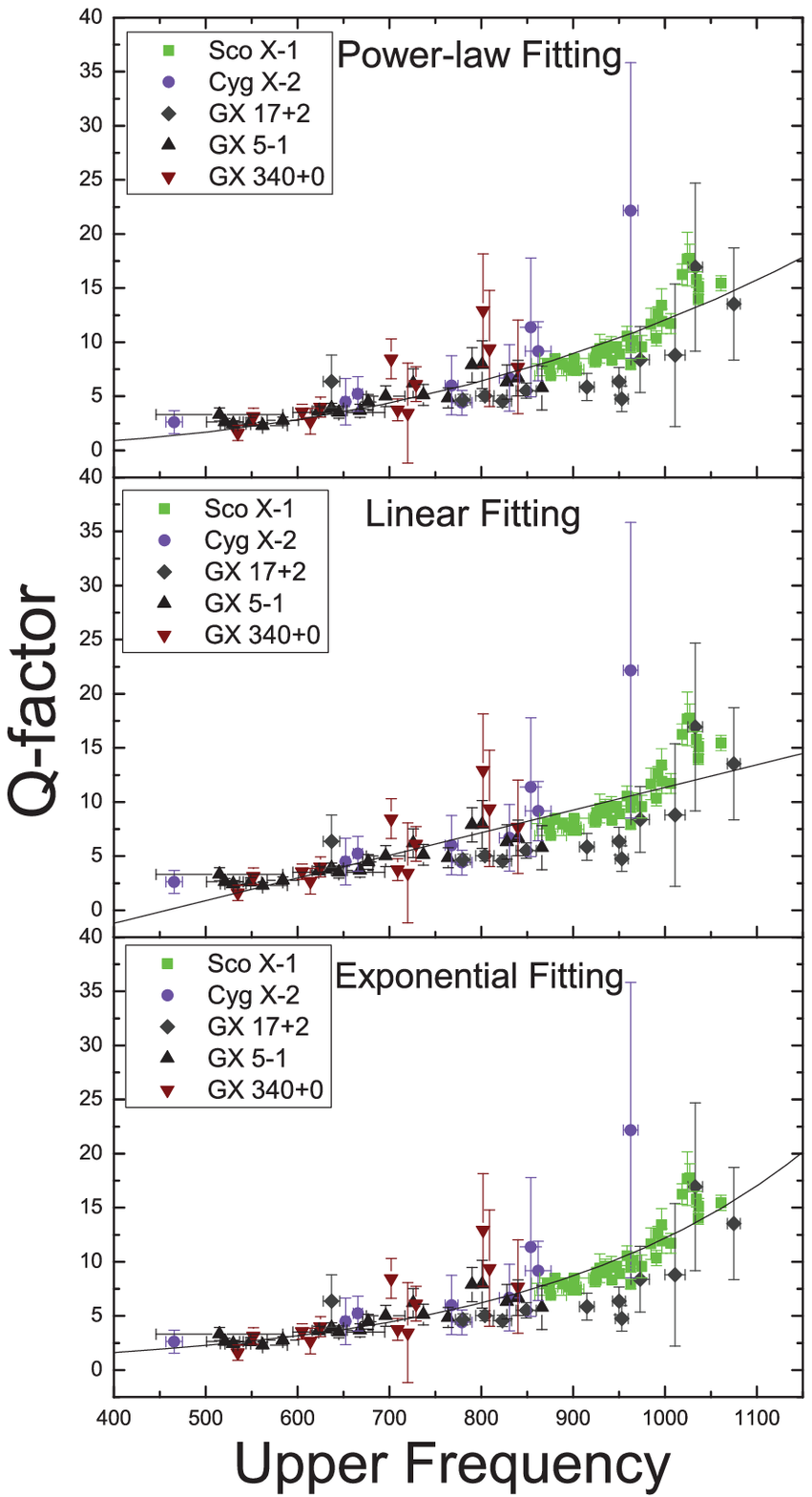}%
\includegraphics[width=5.61cm]{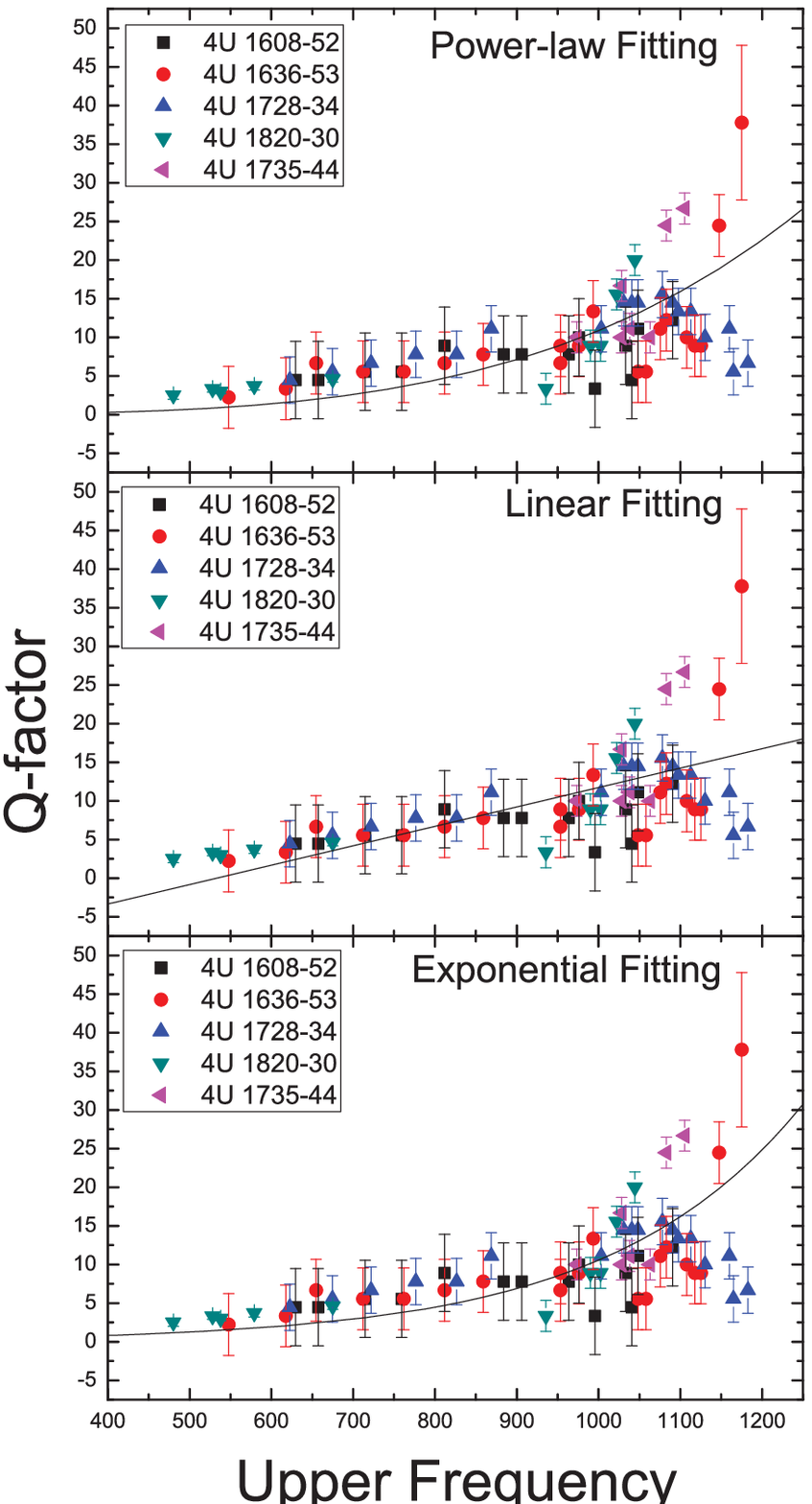}
\caption{The fitting
{\color{red}
plots of Q vs. $\nu$
}
for upper kHz QPOs.
The left panel is for all ten sources. The middle (right) panel is
for five Z (Atoll) sources.
{\color{red}
The three parts from top to bottom
of each panel present the power, linear and exponential fitting,
respectively. The point with different shapes and colors in
the plots represent the different sources as shown in the legends.
The fitting functions and results are shown in table \ref{tab:U}.
}
\label{Fig:upper-fit}}
\end{figure*}

In order to study the $Q_2-\nu_2$
\begin{changed}
relation
\end{changed}
, we
\begin{changed}
use
\end{changed}
linear ($Q = a+b\nu/1000$), power-law ($Q = a(\nu/1000)^b$) and exponential
\begin{changed}
relations
\end{changed}
($Q = a\times
exp(b\nu/1000)$) to
\begin{changed}
fit
\end{changed}
$Q_2-\nu_2$ trajectories for ten sources. It is
noticed that the $Q_2$ values are similar for Z and Atoll sources,
\begin{changed}
which
\end{changed}
increase with $\nu_2$, so we fit the three relations to total
ten sources, Z sources and Atoll sources, respectively (see Fig.
\ref{Fig:upper-fit}). The values of
\begin{changed}
fitting
\end{changed}
parameters ($a$ and $b$) and the
fitting results are listed in table \ref{tab:U}.

From Fig. \ref{Fig:upper-fit}, it is seen that the data points are
dispersive in the left and right panels, while the
\begin{changed}
result in
the
\end{changed}
middle panel is better, especially for power-law and
exponential relations. From table \ref{tab:U}, we notice that
\begin{changed}
the fitting results are not good for these large $\chi^2/d.o.f.$. If
we just forecast a roughly trend based on these results, it can be
seen that
\end{changed}
the power-law and exponential relations fit mildly better
than that of linear relation as a whole. Exponential relation fits
better than power-law and linear relations both for Z sources and
for Atoll sources. The fittings for $Q_2-\nu_2$ of Z sources are
better than that for Atoll sources, with relatively small reduced
$\chi^2$ ($\chi^2/d.o.f. = 4.73, 5.01$ and $5.76$ for exponential,
power-law and linear relations, respectively, see table \ref{tab:U}
for detail). But for five Atoll sources, the reduced $\chi^2$ are
relatively large ($\chi^2/d.o.f.=24.63, 26.10$ and $29.33$ for
exponential, power-law and linear relation, respectively), and the
exponential fitting is also better than the other two relations. If
we put all the sources together and
\begin{changed}
fit
\end{changed}
three relations to
them, we find that the power-law
\begin{changed}
relation
\end{changed}
with a small
$\chi^2/d.o.f.$ (11.75) is better than the other two.

\begin{table*}
\begin{center}
\begin{minipage}{1.5\linewidth}
\caption{The fitting results for quality
{\color{red}
factors
}
of upper kHz QPOs
{\color{red}
correspond to Fig.3
}
.}
\label{tab:U}
\begin{tabular}{ccccc}
\hline \ Function & \ \ \ a \ \ \ & \ \ \ b \ \ \ & \ \ \
$\chi^2/DoF$
 \ \ \ & $R^2$ \\
\hline Total fitting\\
${\color{red} Q_2=a+b\nu_2/1000}$ & ${\color{red} -9\pm2}$ & ${\color{red} 21\pm2}$
& 14.00 & 0.47 \\
${\color{red} Q_2=a(\nu_2/1000)^b}$ & ${\color{red} 10.9\pm0.3}$ & ${\color{red} 2.2\pm0.2}$
& 11.75 & 0.49 \\
${\color{red} Q_2=a\times exp(b\nu_2/1000)}$ & ${\color{red} 0.9\pm0.2}$ & ${\color{red} 2.5\pm0.2}$
& 11.80 & 0.48 \\
\hline Z source \\
${\color{red} Q_2=a+b\nu_2/1000}$ & ${\color{red} -10\pm1}$ & ${\color{red} 21\pm2}$
& 5.76 & 0.64 \\
${\color{red} Q_2=a(\nu_2/1000)^b}$ & ${\color{red} 12\pm0.4}$ & ${\color{red} 2.8\pm0.3}$
& 5.01 & 0.69 \\
${\color{red} Q_2=a\times exp(b\nu_2/1000)}$ & ${\color{red} 0.4\pm0.1}$ & ${\color{red} 3.4\pm0.3}$
& 4.73 & 0.71 \\
\hline Atoll source\\
${\color{red} Q_2=a+b\nu_2/1000}$ & ${\color{red} -13\pm5}$ & ${\color{red} 25\pm5}$
& 29.33 & 0.36 \\
${\color{red} Q_2=a(\nu_2/1000)^b}$ & ${\color{red} 10.9\pm0.9}$ & ${\color{red} 4\pm0.8}$
& 26.10 & 0.43 \\
${\color{red} Q_2=a\times exp(b\nu_2/1000)}$ & ${\color{red}0.1\pm0.1}$ & ${\color{red} 4.3\pm0.8}$
& 24.63 & 0.46 \\
\hline
\end{tabular}
\end{minipage}
\end{center}
\end{table*}

\subsection{Fitting for the $Q_1-\nu_1$ Relations}

\begin{figure*}
\includegraphics[width=8cm]{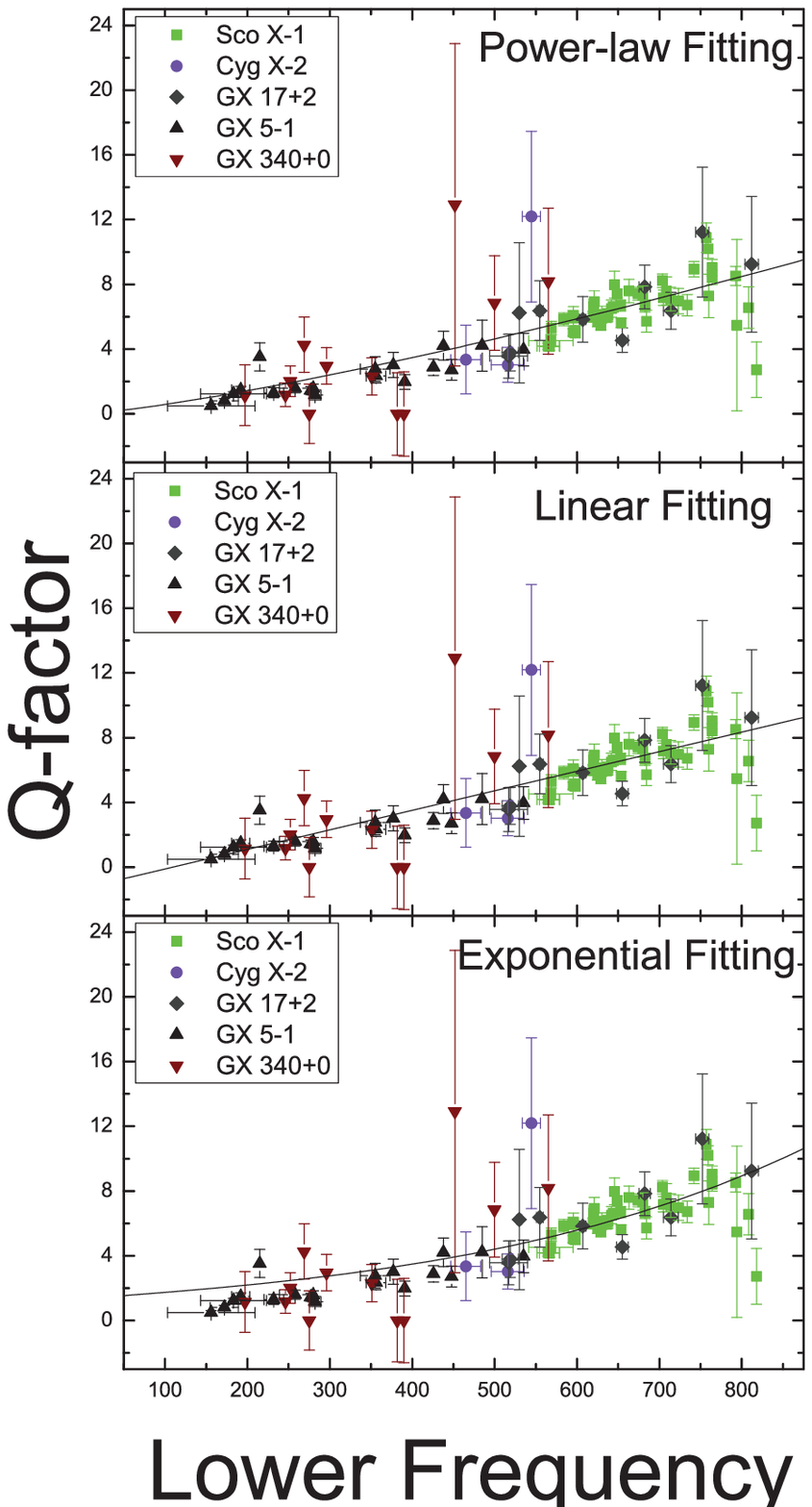}%
\includegraphics[height=14.8cm]{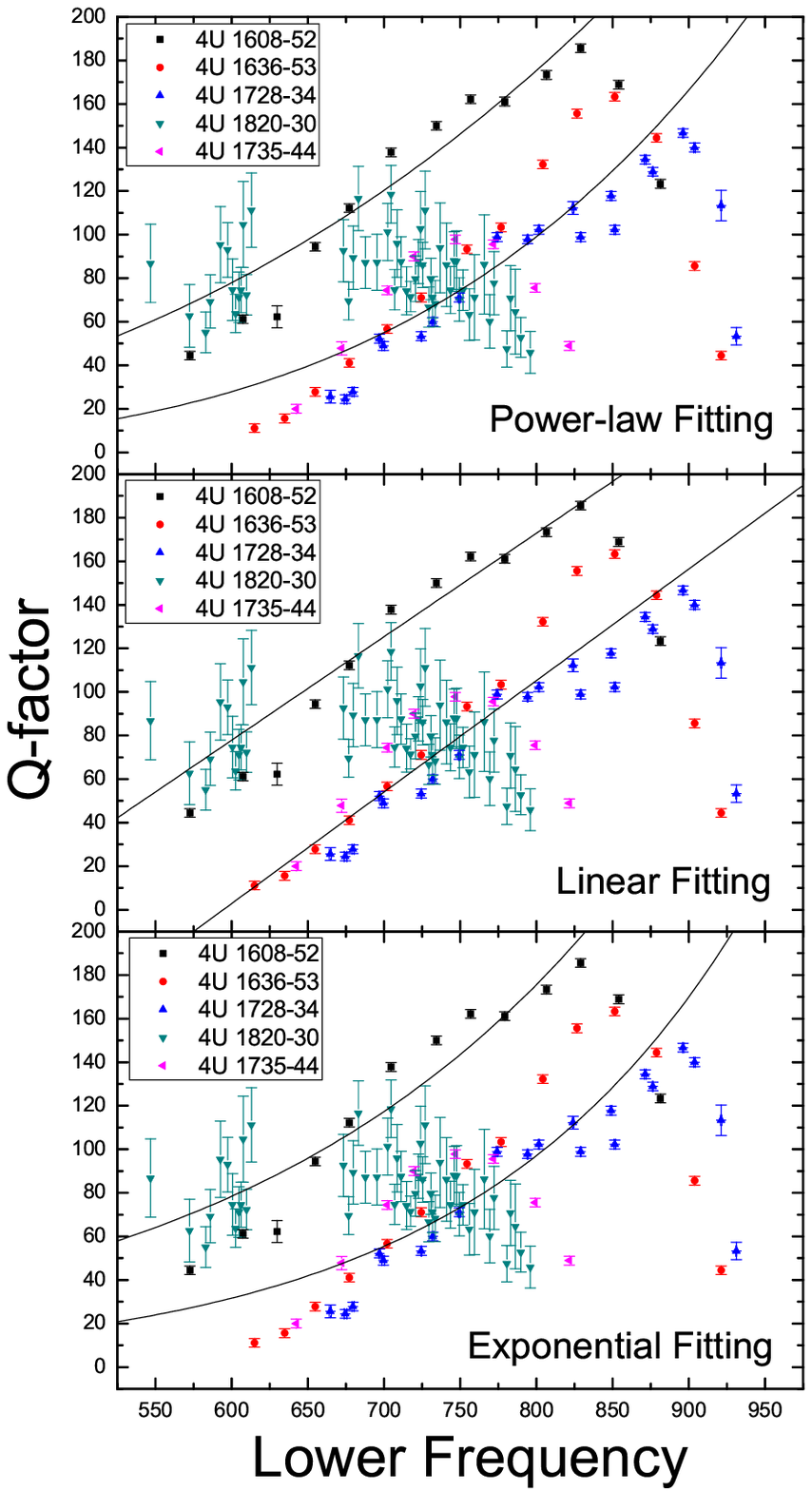}
\caption{The
{\color{red}
similar pattern
}
as Fig. \ref{Fig:upper-fit}, but for lower
kHz QPOs. The left panel is for five Z sources.
The right panel is for five Atoll sources,
{\color{red}
in which we take the sources 4U 1608-52 and
4U 1820-30 (higher quality factors) as a group and the sources
4U 1636-53, 4U1728-34 and 4U 1735-44 (lower quality factors) as the other group, then we fit the data of two
groups, respectively.
}
{\color{red}
The fitting functions and results are shown in Table. \ref{tab:L}.
}
\label{Fig:lower-fit}}
\end{figure*}

Following the same techniques, we
\begin{changed}
investigate
\end{changed}
the $Q_1-\nu_1$ relations for Z and Atoll sources (see
Fig. \ref{Fig:lower-fit}).

The left panel of Fig. \ref{Fig:lower-fit}
\begin{changed}
show the results for Z sources, in which it can be seen that
\end{changed}
almost all points form the rising trend, and we
fit power-law, linear and exponential relations to all the data
points, respectively. The parameters and fitting results are listed
in table \ref{tab:L}. The reduced $\chi^2$ are 3.33, 3.30 and 3.62
for power-law, linear and exponential relation, respectively. It
seems that the linear relation fits better than the other two.

\begin{changed}
The right panel of Fig. \ref{Fig:lower-fit}
show the results for Atoll sources, in which it can be seen that the
\end{changed}
$Q_1-\nu_1$
\begin{changed}
distribution of
\end{changed}
five Atoll sources display two
rising parts and two dropping parts (see
\begin{changed}
the right panel of
\end{changed}
Fig. \ref{Fig:lower-fit}
for detail).
\begin{changed}
We take the sources 4U 1608-52 and 4U 1820-30, which have large $Q_1$ values, as a group.
While the sources 4U 1636-53, 4U 1728-34 and 4U 1735-44, which have small $Q_1$ values, as the other group.
Then we fit the data of the two groups respectively, the fitting results are in table \ref{tab:L}.
Compared with the fitting result of $Q_2-\nu_2$,
\end{changed}
the reduced $\chi^2$ is very large
(see table \ref{tab:L} for detail), but also with large $R^2$. It is
found that the linear relation fits better than the other two for
the curve with relatively low $Q_1$ values (formed by 4U 1636-53, 4U
1728-34 and 4U 1735-44), and the power-law relation fits better to
the curve with high $Q_1$ values (formed by 4U 1608-52 and 4U
1820-30).

So far, there is no appropriate model which can explain the
mechanism of lower kHz QPOs, as well as that
\begin{changed}
of
\end{changed}
the lower quality
factors. However, it is
\begin{changed}
claimed
\end{changed}
that the lower kHz QPOs arise in the
innermost region of accretion disk (van der Klis 2006). The quality
factors relate to the frequency drift in the inner disk (Barret,
Olive \& Miller 2006; Wang et al. 2011). So we claim that the
diverge of the data points between different sources
\begin{changed}
may result
\end{changed}
from
\begin{changed}
different physical environment
\end{changed}
of inner disk
\begin{changed}
in
\end{changed}
different sources. In
addition, the occurrence of drops in $Q_1-\nu_1$ plots for Atoll
sources may arise from the existence of the inner boundary of
accretion disk (Barret, Olive \& Miller 2006). Due to the different
mechanism of the drops of $Q_1-\nu_1$ tracks for Atoll sources, we
just fit the rising branches of the $Q_1-\nu_1$ track and do not
consider the dropping parts with
\begin{changed}
the
\end{changed}
same techniques.

\begin{table*}
\begin{center}
\begin{minipage}{1.5\linewidth}
\caption{The fitting results for quality {\bf factors} of {\bf lower} kHz QPOs {\bf correspond to Fig.4}.}
\label{tab:L}
\begin{tabular}{ccccc}
\hline \ Function & \ \ \ a \ \ \ & \ \ \ b \ \ \ & \ \ \
$\chi^2/DoF$
\ \ \ & $R^2$ \\
\hline Z source \\
${\color{red} Q_1=a+b\nu_1/1000}$ & ${\color{red} -1.3\pm0.6}$ & ${\color{red} 12\pm1}$
& 3.30 & 0.59 \\
${\color{red} Q_1=a(\nu_1/1000)^b}$ & ${\color{red} 11.3\pm0.9}$ & ${\color{red} 1.3\pm0.2}$
& 3.33 & 0.59 \\
${\color{red} Q_1=a\times exp(b\nu_1/1000)}$ & ${\color{red} 1.4\pm0.3}$ & ${\color{red} 2.4\pm0.3}$
& 3.62 & 0.56 \\
\hline Atoll source with lower $Q_1$\\
${\color{red} Q_1=a+b\nu_1/1000}$ & ${\color{red} -300\pm30}$ & ${\color{red} 500\pm30}$
& 241.70 & 0.87 \\
${\color{red} Q_1=a(\nu_1/1000)^b}$ & ${\color{red} 260\pm30}$ & ${\color{red} 4.4\pm0.4}$
& 343.75 & 0.82 \\
${\color{red} Q_1=a\times exp(b\nu_1/1000)}$ & ${\color{red} 1.1\pm0.5}$ & ${\color{red} 5.6\pm0.6}$
& 380.99 & 0.80 \\
\hline Atoll source with higher $Q_1$\\
${\color{red} Q_1=a+b\nu_1/1000}$ & ${\color{red} -200\pm30}$ & ${\color{red} 480\pm40}$
& 270.52 & 0.85 \\
${\color{red} Q_1=a(\nu_1/1000)^b}$ & ${\color{red} 330\pm30}$ & ${\color{red} 2.8\pm0.2}$
& 267.59 & 0.85 \\
${\color{red} Q_1=a\times exp(b\nu_1/1000)}$ & ${\color{red} 7\pm2}$ & ${\color{red} 4\pm0.3}$
& 279.81 & 0.84\\
\hline
\end{tabular}
\end{minipage}
\end{center}
\end{table*}

\section{ Discussions and Summary}
In this work, we study the distribution of
\begin{changed}
the
\end{changed}
quality factors
and the relation between
\begin{changed}
the
\end{changed}
quality
\begin{changed}
factor and peak
frequency of kHz QPOs
\end{changed}
. We notice that there are differences in the
distribution of
\begin{changed}
the
\end{changed}
quality factors between lower and upper
kHz QPOs, as well as Z and Atoll sources. By fitting the power-law,
linear and exponential relation to the $Q_\nu$ tracks, we
investigate the $Q_\nu$ relations for upper and lower kHz QPOs.

(1). The $Q_2$ values are low ($Q_2 < 18$ in general, with a few
exceptions) for both Z and Atoll sources. The $Q_1$ values for five
Z sources are very low, $Q_{1z} = 2-14$. However, the $Q_1$ values
for five Atoll sources are very high, and the maximum $Q_{1Atoll}$
is up to 200 (4U 1608-52).

(2). On the whole, the ranges of Q for five Atoll sources are wider
than that for five Z sources both for upper and lower Q. According
to the idea that the twin kHz QPOs come from the inner region of
accretion disk, the instabilities which can lead to the radial drift
and frequency drift in this region for Atoll sources are stronger
than that for Z sources (Wang et al. 2011).

(3). From Fig. \ref{Fig:lower-his}, it seems that the range of $Q_1$
for Atoll sources is about 10 times wider than that for Z sources.
In addition, the high luminosity Z sources exhibit luminosity close
to critical Eddington luminosity $L_{Edd}$, and the range of their
luminosity is $0.5-1 L_{Edd}$. But Atoll sources present low
luminosity $0.001-0.2 L_{Edd}$, 100 times lower than that for Z
sources (Hasinger and van der Klis 1989; Hasingger 1990, see van der
Klis 2006 for a review). Therefore, we expect a relation of $Q_1
\sim L^{-1/2}$ between the luminosity and quality factors for lower
kHz QPOs to be investigated and proven in future.

(4).
\begin{changed}
The fitting results of $Q_2~vs. \nu_2$ can not confirm the
actual relation between $Q_2$ and $\nu_2$ because of the large
$\chi^2/d.o.f.$. We consider that it may result from the small
sample with the  relative big errors. There is an obvious
trend that
\end{changed}
$Q_2$ increases with $\nu_2$ for both Z and Atoll
sources,
\begin{changed}
and
\end{changed}
the power-law and exponential relations fit
better than the linear relation.

(5).
\begin{changed}
The fitting results of $Q_1~vs. \nu_1$ also cannot be
considered as perfect. Then, an obvious trend is that
\end{changed}
the $Q_1$
increase with $\nu_1$ in the case of five Z sources, and the linear
relation fits better than the other two.

(6). In the $Q_1-\nu_1$ diagrams for Atoll sources, all five sources
present drops at a maximum $Q_1$.
\begin{changed}
It is considered as the
results of innermost boundary (ISCO) of accretion disk, firstly
proposed by Barret and coauthors (see, Barret et al. 2005b, 2005c,
2006, 2007)
\end{changed}
. The special mechanism of the drops are expected to be
investigated further. Here, we just fit the three relations to the
rising parts.)

(7).
\begin{changed}
As for the relation between $Q_1$ and $\nu_1$ in Atoll
sources, it can be seen in the right panel of Fig.
\ref{Fig:lower-fit} that,  if only the rising part of $Q_1 - \nu_1$
plot is considered, the five Atoll sources form two curve-like
tracks, but it is not enough  obvious to indicate a clear
correlation. We claim that this phenomenon arises from the special
physics in the inner disk region for each source, where stellar
mass, magnetic field and accretion rate may be the roles of
producing them. For the fitting, we fit the three relations to each
rising curve.  This tentative fitting is not good enough to show a
clear correlation although the $\chi^2/d.o.f.$
\end{changed}
of linear
\begin{changed}
fitting
\end{changed}
is better than the other two for the curve with relatively
low $Q_1$ values (formed by 4U 1636-53, 4U 1728-34 and 4U 1735-44),
and
\begin{changed}
the $\chi^2/d.o.f.$ of
\end{changed}
the power-law
\begin{changed}
fitting
\end{changed}
is
better to the curve with high $Q_1$ values (formed by 4U 1608-52 and
4U 1820-30).

\acknowledgements We acknowledge M. M\'{e}ndez and D. Barret for
providing the data. This work is supported by the National Natural
Science Foundation of China (NSFC 10773017, NSFC 10773034, NSFC
10778716, NSFC 11173024), the National Basic Research Program of
China (2009CB824800, 2012CB821800), NSC 99-2112-M-007-017-MY3, and
the Fundamental Research Funds for the Central Universities. We are
very grateful for the anonymous referee for critic  comments, which
has changed the quality of the paper.


\begin{thebibliography}{}

\bibitem{}
Abramowicz, M. A., Karas, V., Kluzniak, W., Lee, W. H., Rebusco, P.:
2003, PASJ~55, 467

\bibitem[Barret et al.(2008)]{Barret08}
Barret, D., Boutelier, M., \& Miller, M.~C.: 2008, MNRAS~384, 1519

\bibitem[Barret (2005a)]{}
Barret, D., Klu$\acute{z}$niak, W., Olive, J. F., Paltani, S.,
Skinner, G. K.: 2005a, MNRAS~357, 1288

\bibitem[Barret (2005b)]{}
Barret, D., Olive, J. F., Miller, M. C.: 2005b, MNRAS~361, 855

\bibitem[Barret (2005c)]{}
Barret, D., Olive, J. F., Miller, M. C.: 2005c, Astron. Nachr.~326,
808

\bibitem[Barret (2006)]{}
Barret, D., Olive, J. F., Miller, M. C.: 2006, MNRAS~370, 1140

\bibitem[Barret et al.(2007)]{2007MNRAS.376.1139B}
Barret, D., Olive, J. F., Miller, M. C.: 2007, MNRAS~376, 1139

\bibitem{}
Belloni, T., Mendez, M., Homan, J.: 2005, A\&A~437, 209

\bibitem{}
Belloni, T., M\'{e}ndez, M. \& Homan, J.: 2007, MNRAS~376, 1133

\bibitem{}
Belloni, T., Psaltis, D., van der Klis, M.: 2002, ApJ~572, 392

\bibitem[Boutelier et al.(2009)]{2009MNRAS.399.1901B} Boutelier, M.,
Barret, D., \& Miller, M.~C.: 2009, MNRAS~399, 1901

\bibitem[Boutelier et al.(2010)]{2010MNRAS.401.1290B} Boutelier, M.,
Barret, D., Lin, Y., T{\"o}r{\"o}k, G.: 2010, MNRAS~401, 1290

\bibitem{Hasinger1989}
Hasinger, G., \& van der Klis, M.: 1989, A\&A~225, 79

\bibitem{Hasinger1990}
Hasinger, G.: 1990, RvMA~3, 60


\bibitem{}
Kluzniak, W., Abramowicz, M. A.: 2001, Acta Physica Polonica B~32,
3605

\bibitem[mendez (2006)]{}
M$\acute{e}$ndez, M.: 2006, MNRAS~371, 1925

\bibitem{}
M$\acute{e}$ndez, M., van der Klis, M.: 1999, ApJ~517, L51

\bibitem{men2000}
M\'endez, M., \& van der Klis, M. 2000, MNRAS~318, 938

\bibitem{}
M\'endez, M., van der Klis, M., \& Ford, E.C. 2001, \apj~561, 1016

\bibitem[Miller (1998)]{}
Miller, M. C., Lamb ,F. K., Psaltis, D.: 1998, ApJ~508, 791

\bibitem{}
Psaltis, D. et al. 1998, ApJ~501, L95

\bibitem{}
Psaltis, D. et al. 1999a, ApJ~520, 763

\bibitem{}
Psaltis, D., Belloni, T., van der Klis, M.: 1999b, ApJ~520, 262

\bibitem{Stella1998}
Stella, L. \& Vietri, M.: 1998, ApJ~492, L59

\bibitem{Stella1999}
Stella, L. \& Vietri, M.: 1999, PRL~82, 17

\bibitem{SVM1999}
Stella, L., Vietri, M., Morsink, S. M.: 1999, ApJ~524, L63

\bibitem{}
Titarchuk, L. \& Wood, K.: 2002, APJ~577, L23

\bibitem[Torok (2009)]{} T$\ddot{o}$r$\ddot{o}$k, G.: 2009, A\&A~497, 661

\bibitem[van der Klis (2000)]{}
van der Klis, M.: 2000, ARA\&A~38, 717 

\bibitem[van der Klis (2006)]{}
van der Klis, M.: 2006, A review of rapid X-ray variability in X-ray
binaries in Compact stellar X-ray sources, W.H.G. Lewin \& M. van
der Klis (eds. ), Cambridge University Press, p. 39;
(astro-ph/0410551).

\bibitem{}
van der Klis, M.: 2008, AIPC.~1068, 163

\bibitem{}
Wang, J., Zhang C. M., Zhao Y. H., Lin Y. F., Yin H. X., Song L. M.:
2011, A\&A~528, 126

\bibitem{}
Yu W. F., van der Klis M. 2002, ApJ~567, 67

\bibitem{}
Yu W. F., van der Klis M., Jonker P. G. 2001, ApJ~559, 29

\bibitem[zhang (2004)]{}
Zhang, C. M.: 2004, A\&A~423, 401

\bibitem{}
Zhang, C. M., Yin, H. X., Zhao, Y. H., Zhang, F. \& Song, L. M.:
2006, MNRAS~366, 1373

\end{thebibliography}
\end{document}